\documentclass[%
reprint,
superscriptaddress,
showkeys,
amsmath,amssymb,
prl,
longbibliography,
floatfix,
nobalancelastpage
]{revtex4-2}

\usepackage{defs}

\begin{document}
\title{Spin-Polarized Photoelectrons in the Vicinity of Spectral Features}

\author{Stefanos~Carlström\,\orcidlink{0000-0002-1230-4496}}%
\email{stefanos@mbi-berlin.de}
\affiliation{Max-Born-Institut, Max-Born-Straße 2A, 12489 Berlin, Germany}
\affiliation{Department of Physics, Lund University, Box 118, SE-221 00 Lund, Sweden}

\author{Rezvan~Tahouri\,\orcidlink{0009-0002-0839-3606}}%
\affiliation{Department of Physics, Lund University, Box 118, SE-221 00 Lund, Sweden}

\author{Asimina~Papoulia\,\orcidlink{0000-0002-3658-3522}}%
\affiliation{Department of Physics, Lund University, Box 118, SE-221 00 Lund, Sweden}

\author{Jan~Marcus~Dahlström\,\orcidlink{0000-0002-5274-1009}}%
\affiliation{Department of Physics, Lund University, Box 118, SE-221 00 Lund, Sweden}

\author{Misha~Yu~Ivanov\,\orcidlink{0000-0002-8817-2469}}%
\affiliation{Max-Born-Institut, Max-Born-Straße 2A, 12489 Berlin, Germany}
\affiliation{Institut für Physik, Humboldt-Universität zu Berlin, Newtonstraße 15, 12487 Berlin, Germany}
\affiliation{Technion, Israel Institute of Technology, 3200003 Haifa, Israel}

\author{Olga~Smirnova\,\orcidlink{0000-0002-7746-5733}}%
\affiliation{Max-Born-Institut, Max-Born-Straße 2A, 12489 Berlin, Germany}
\affiliation{Technische Universität Berlin, Ernst-Ruska-Gebäude, Hardenbergstraße 36A, 10623 Berlin, Germany}
\affiliation{Technion, Israel Institute of Technology, 3200003 Haifa, Israel}

\author{Serguei~Patchkovskii}%
\affiliation{Max-Born-Institut, Max-Born-Straße 2A, 12489 Berlin, Germany}

\date{\today}

\begin{abstract}
  It has been shown by \textcite{Fano1969} that photoionization of a
  cæsium atom by a laser pulse tuned to the vicinity of a Cooper
  minimum generates spin-polarized electrons. Here we show that while
  photoionization of rare gases does not provide large spin
  polarization in the vicinity of the Cooper minimum, the Fano
  resonances yield much higher overall spin polarization
  (\(\ge\SI{40}{\percent}\)).  The spin polarization increases in
  angle-resolved photoelectron spectra, and reaches \SI{100}{\percent}
  when measured in coincidence with the photoion. We provide a general
  framework for achieving spin polarization in photoionization
  irrespective of the ionization regime.
\end{abstract}


\maketitle



The efficient generation of spin-polarized electrons is interesting
for a multitude of reasons, including the table-top, laser-driven
particle accelerators \cite{Nie2021,Nie2022a}, and more recently, a
proposed Bell test of quantum entanglement \cite{Ruberti2024}. To
generate spin-polarized photoelectrons from multi-electron atoms,
several prerequisites have to be satisfied \cite{Barth2013}. First,
spin--orbit interaction is needed to entangle spin and angular degrees
of freedom. Whenever the photoionization amplitudes exhibit
sensitivity to orbital angular momentum, we may thus expect
selectivity with respect to spin. Second, a mechanism is needed to
break the balance in ionization to continua with different
combinations of spin \(S\) and orbital \(L\) angular momenta. As the
ionizing laser field couples to the orbital angular momentum rather
than spin, selectivity in ionization with respect to different orbital
angular momenta is also needed.

These general principles are already implicit in the pioneering work
of \textcite{Fano1969}. He suggested to generate spin-polarized
photoelectrons by taking advantage of the Cooper minimum (CM) found in
photoionization of cæsium. Starting from a
\(\conf{6s\;\term{1}{S}[0]}\) ground state, \(\sigma^+\) circularly
polarized light will generate \(\conf{p}_+\) photoelectrons. For
spin-up electrons, this will predominantly lead to the \(J=3/2\)
ionization channel, and to the \(J=1/2\) ionization channel for
spin-down electrons. In these channels, the Cooper minima appear
slightly shifted energetically with respect to each other, allowing us
to select the desired ionization channel, and thus dominant spin
contribution, by tuning the photon energy. From the conservation of
angular momentum perspective, the spin--orbit coupling transfers
spin-angular momentum of the photon to the emitted electron. This
prediction was experimentally confirmed by \textcite{Heinzmann1970},
with a spin polarization \(\ge\SI{80}{\percent}\).

In the case of few-photon ionization, using intermediate resonances
with bound states may allow one to enhance a desired \(J\)-dependent
ionization pathway to achieve high spin polarization
\cite{Lambropoulos1973}. In the strong field ionization regime known
as nonadiabatic optical tunneling \cite{Yudin2001}, electron tunneling
through the barrier created by circularly polarized field and the core
potential is also selective with respect to the angular momentum of
the tunneling electron, leading to spin polarization predicted in
\cite{Barth2013,Barth2014,Barth2014a,Nie2021} and experimentally
confirmed in \cite{Hartung2016,Trabert2018}. The required selectivity
in ionization to continua with different \(J\) arises from the
exponential sensitivity of the tunneling process to the
\(J\)-dependent ionization potential.

Here we return to the one-photon regime originally studied by Fano,
but focus on the role of the continuum resonances bearing his name. We
use \emph{ab initio} calculations to describe photoionization of argon
and xenon in a broad energy range of photon energies (up to
\SI{250}{\electronvolt}). This includes the autoionizing resonances
leading up to the \(n\conf{s}\), \(J=1/2\) (argon and xenon), and
\(4\conf{d}\), \(J=3/2\) and \(J=5/2\) (xenon) thresholds (double
ionization, which would broaden some Fano lines \cite{Svensson1976},
is not included).

In view of the general scheme outlined above, these resonances offer
important advantages. First, they exhibit a strong angular
dependence. Second, their minima are clear and sharp, in contrast to
shallow spectral features such as the CM or shape resonances. This
allows for high angle- and energy-dependent selectivity of the
ionization channel. Altogether, this leads to excellent spin
selectivity, as demonstrated below. Note that one can also select a
specific ionization channel (for noble gases \(J=3/2\)), by choosing a
spectrally narrow photon energy below any of the higher-lying
ionization channels
\cite{Dill1973,Lee1974,Fano1975,Geiger1977,Johnson1980,Heinzmann1980I,Heinzmann1980II}.

The qualitative effect of the CM on the photoelectron spin identified
by Fano is not unique to cæsium, but the quantitative strength is. We
find that in argon and xenon, the rare gas atoms that exhibit
well-known Cooper minima, the associated spin polarization is only
about \SI{2}{\percent}. Therefore, it is desirable to find an
alternative path to generate spin-polarized photoelectrons from these
targets. Indeed, we find substantially higher spin polarization in the
vicinity of the Fano resonances, in excess of
\SI{40}{\percent}. Furthermore, if measured in coincidence with the
ion, angularly resolved photoelectron spectra show spin polarization
approaching \SI{100}{\percent}. We complement our numerical
calculations with a simple analytical model describing the underlying
physical effect.

\begin{figure}
  \centering
  \includegraphics[trim=30cm 30cm 30cm 30cm]{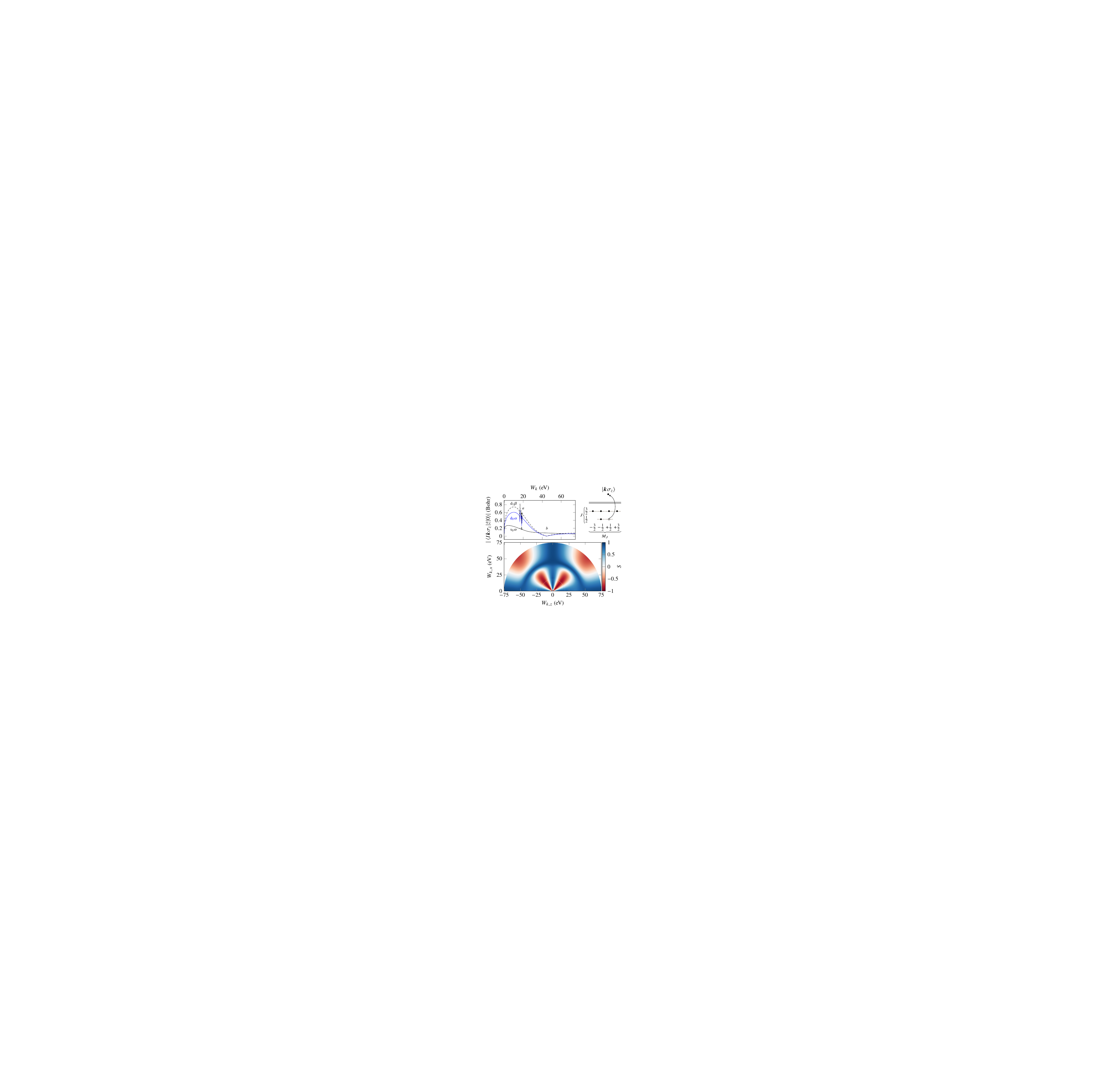}
  \caption{Argon photoionization dipole matrix elements (\emph{top
      left}) measured in coincidence with the ion in the
    \archannelb{}, \(M_J=1/2\) state (\emph{top right}), for the LP
    dipole moment. The different lines correspond to contributions
    from different partial waves and spins (solid spin-up and dashed
    spin-down). The Fano resonances leading to the \archannelc{}
    threshold are indicated by \emph{a} and the location of the CM by
    \emph{b}. Shown is also the resultant spin polarization
    (\emph{bottom}), where in particular the CM manifests itself as a
    circle around \(\kineng=\SI{50}{\electronvolt}\) where the emitted
    photoelectron is \SI{100}{\percent} spin-up.}
  \label{fig:argon-channel-dmes}
\end{figure}

The spin polarization along the quantization axis \(+z\), for
one-photon ionization into a state with photoelectron momentum
\(\vec{k}\), is given by the photoionization matrix elements between
the ground state and the final scattering state \cite{Fano1969}:
\begin{equation}
  \label{eqn:spin-polarization-matrix-elements}
  S_z^I \defd
  \frac{P_{I\spinup}-P_{I\spindown}}{P_{I\spinup}+P_{I\spindown}}
  \approx
  \frac{\abs*{\matrixel*{I\vec{k}_\spinup}{\operator{d}}{0}}^2-
    \abs*{\matrixel*{I\vec{k}_\spindown}{\operator{d}}{0}}^2}
  {\abs*{\matrixel*{I\vec{k}_\spinup}{\operator{d}}{0}}^2+
    \abs*{\matrixel*{I\vec{k}_\spindown}{\operator{d}}{0}}^2}.
\end{equation}
Here, \(P_{I\spinup,\spindown}\) are the probabilities of measuring an
\(\spinup\) (spin-up) or \(\spindown\) (spin-down) electron,
respectively, \(\ket{I}\) denotes the final state of the ion
(configuration \(n\ell_J^{-1}\) and magnetic quantum number \(M_J\)),
\(\ket{0}\) the ground state of the atom. Finally,
\(\operator{d}=\operator{z}\) for linearly polarized (LP) light,
\(\operator{d}=(\operator{x}+\im\xi\operator{y})/\sqrt{2}\) with
\(\xi=1\) referred to as right-hand circularly polarized (RCP) light,
and \(\xi=-1\) referred to as left-hand circularly polarized (LCP)
light. Summing over the final ion states, yields the total (overall)
spin polarization:
\begin{equation}
  \label{eqn:total-spin-polarization-matrix-elements}
  S_z^\textrm{(tot)} \approx
  \frac{\sum_I\abs*{\matrixel*{I\vec{k}_\spinup}{\operator{d}}{0}}^2-
    \abs*{\matrixel*{I\vec{k}_\spindown}{\operator{d}}{0}}^2}
  {\sum_I\abs*{\matrixel*{I\vec{k}_\spinup}{\operator{d}}{0}}^2+
    \abs*{\matrixel*{I\vec{k}_\spindown}{\operator{d}}{0}}^2}.
\end{equation}

Conservation of angular momentum dictates that photoionization by LP
light (which carries no spin-angular momentum) cannot generate net
spin-angular momentum upon photoionization of a spin-0 target, such as
the noble gas atoms initially in their ground state:
\(S_z^\textrm{(tot)}=0\). We note in passing, however, that
ion-channel resolved spin polarization may be non-zero even in this
case \cite{Carlstroem2023linspinpol}, provided the photoelectron is
measured in coincidence with the ion, including measuring the \(M_J\)
quantum number, thereby establishing the chiral reference frame (the
so-called \enquote{chiral observer} \cite{Ayuso2022}). This is a
formidable task. Another viewpoint is that for spin polarization to
occur, we need to be able to trace the photoelectron back to the
\(M_J\) it originated from, which for LP is only possible if the ion
is simultaneously measured. For circular polarization, this selection
of \(M_J\) happens naturally \cite{Barth2013}.

To compute the spin polarization, we thus need the dipole matrix
elements from the ground state (or in general any bound state) to the
scattering states, resolved on ion channel and photoelectron momentum
and spin. These matrix elements are computed using the multi-channel
extension of the infinite-time surface flux technique
\cite{Morales2016-isurf} described in \cite{Carlstroem2022tdcisI}:
\begin{equation}
  \label{eqn:dipole-matrix-element}
  \matrixel*{I\vec{k}_{\spin}}{\operator{d}}{0} =
  \sqrt{N}
  \bra{\vec{k}_{\spin}}
  \comm*{\scatteringhamiltonian}{\heaviside(r_s)}
  \matrixell*{I}{(\Hamiltonian_0-\epsilon)^{-1}\operator{d}}{0},
\end{equation}
where \(\scatteringhamiltonian\) is the asymptotic Hamiltonian obeyed
by the wavefunction and the scattering state in the region beyond the
matching surface \(r_s\), \(\epsilon=E_I + k^2/2\) is the total energy of the
scattering state, and the resolvent
\((\Hamiltonian_0-\epsilon)^{-1}\), intimately related to inverse iterations,
formally propagates the wavefunction from the time of the perturbation
to the time of detection. The full derivation is given in
\cite{Carlstroem2022tdcisI}.

The advantage of this expression, compared to the volume integral of
the left-hand side of Eq.~\eqref{eqn:dipole-matrix-element}, is
that we can compute the action of the resolvent on the
dipole-disturbed ground state in linear time. The remaining surface
integral reduces to linear combinations of the wavefunction amplitudes
and derivatives at \(r=r_s\) with the corresponding amplitudes and
derivatives of the scattering states, since they are expanded in terms
of spherical harmonics. The scattering states are accurately evaluated
in terms of Coulomb waves \cite{Barnett1996}. The electronic structure
is described in the Supplementary Information (SI) \citetheSI{}.

\begin{figure}
  \centering
  \includegraphics[trim=30cm 30cm 30cm 30cm]{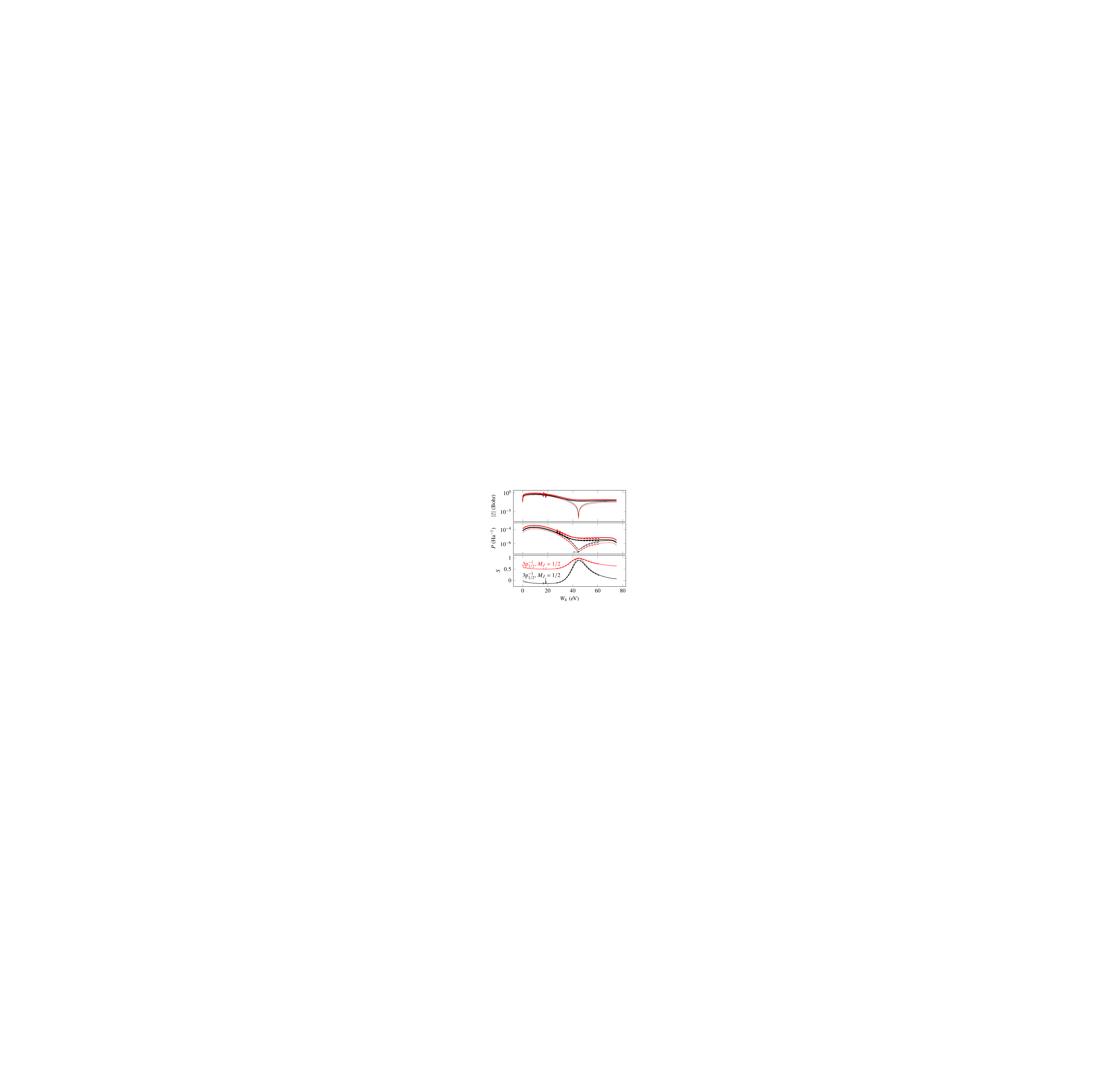}
  \caption{Argon angle-integrated spin-resolved populations and spin
    polarization when measured in coincidence with the ion in the
    \archannelb{}, \(M_J=1/2\) state (black lines) or in the
    \archannela{}, \(M_J=1/2\) state (red lines), for LP. \emph{Top
      panel}: squared dipole matrix elements computed using the PT
    Eq.~\eqref{eqn:dipole-matrix-element}, thick lines
    \(\spinup\), thin lines \(\spindown\)); \emph{middle panel}: solid
    lines spin-resolved populations after convolving the dipole matrix
    elements with a Gaussian spectrum of \SI{6}{\electronvolt}
    bandwidth, dashed lines using time-propagation of the RTDCIS (see
    main text); \emph{bottom panel}: spin polarization, dotted lines
    computed using the PT results from the top panel, solid lines
    convolved PT results, dashed lines using RTDCIS. The RTDCIS
    results have been blueshifted by \SI{4.25}{\electronvolt} to align
    the CM between the calculations.}
  \label{fig:argon-channel-angle-integrated-spin-pol}
\end{figure}

As an illustrative example, we show in
Fig.~\ref{fig:argon-channel-dmes} the dipole matrix elements for
ionization using LP into the \archannelb{}, \(M_J=1/2\) channel of
argon, calculated according to Eq.~\eqref{eqn:dipole-matrix-element}
(see the SI \citetheSI{} for the very similar dipole moments for LCP
and RCP). The Cooper minimum around \(\kineng=\SI{45}{\electronvolt}\)
is clearly visible in that the matrix elements for ionization to the
\conf{d} symmetry vanishes there \cite{Cooper1962}, leaving only the
\conf{s} contribution, which is spin pure. Precisely at the CM, we
thus expect \SI{100}{\percent} spin polarization of the photoelectron,
when measured in coincidence with the photoion. This is indeed the
case, as we see in Fig.~\ref{fig:argon-channel-dmes}, and this remains
true if we angularly integrate the spin-resolved photoelectron
distribution before computing the spin polarization; see
Fig.~\ref{fig:argon-channel-angle-integrated-spin-pol}. For an
independent confirmation based on the solution of the Dirac equation,
we compute the spin-resolved time-dependent ionization fluxes using
the relativistic time-dependent configuration-interaction singles
(RTDCIS) \cite{Zapata2022,Tahouri2024}, for an ionizing pulse of
finite duration (\SI{300}{\atto\second} \(\implies\) \(\sim\)
\SI{6}{\electronvolt} bandwidth;
\SI{1}{\tera\watt\per\centi\meter\squared}) and a set of photon
energies around the CM; see the SI \citetheSI{} for a description on
how the spin-resolved populations are extracted from the fluxes. We
find the predicted spin polarizations to be in good agreement, when
considering that convolving with the finite pulse smooths out the
infinite-time response computed by perturbation theory and
Eq.~\eqref{eqn:spin-polarization-matrix-elements}.

\begin{figure}
  \centering
  \includegraphics[trim=30cm 30cm 30cm 30cm]{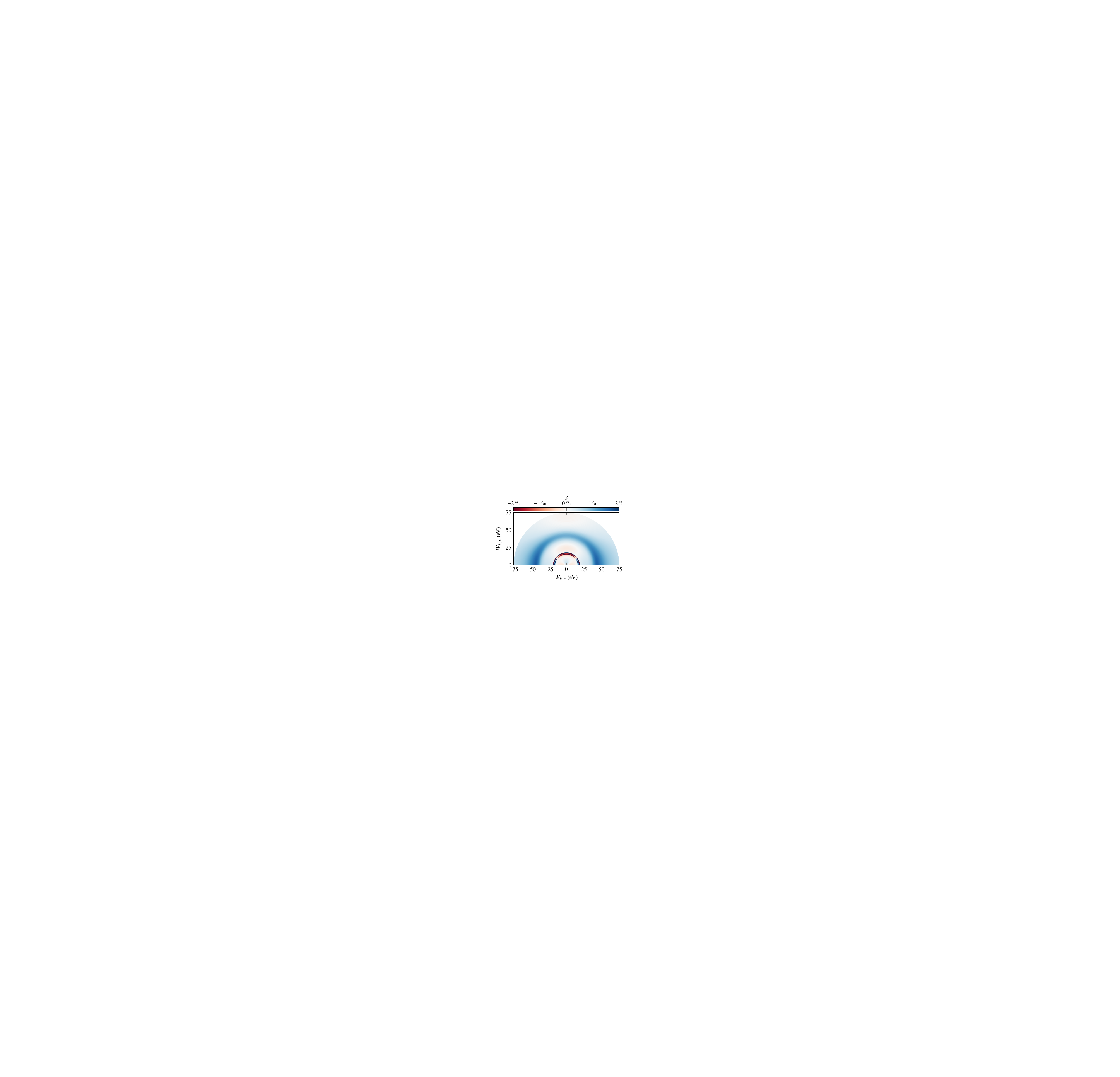}
  \caption{Angle-resolved argon total spin polarization, i.e.\
    obtained by summing over all ion channels, when ionizing with LCP
    light (RCP will give exactly the opposite result, and LP will
    yield exactly zero spin polarization). Note that the colour scale
    saturates at \SI{\pm2}{\percent}, which is the approximate maximum
    spin polarization obtained in the vicinity of the CM, whereas in
    the region of the Fano resonances, the spin polarization exceeds
    \SI{40}{\percent}.}
  \label{fig:argon-total-spin-pol}
\end{figure}

\begin{figure}
  \centering
  \includegraphics[trim=30cm 30cm 30cm 30cm]{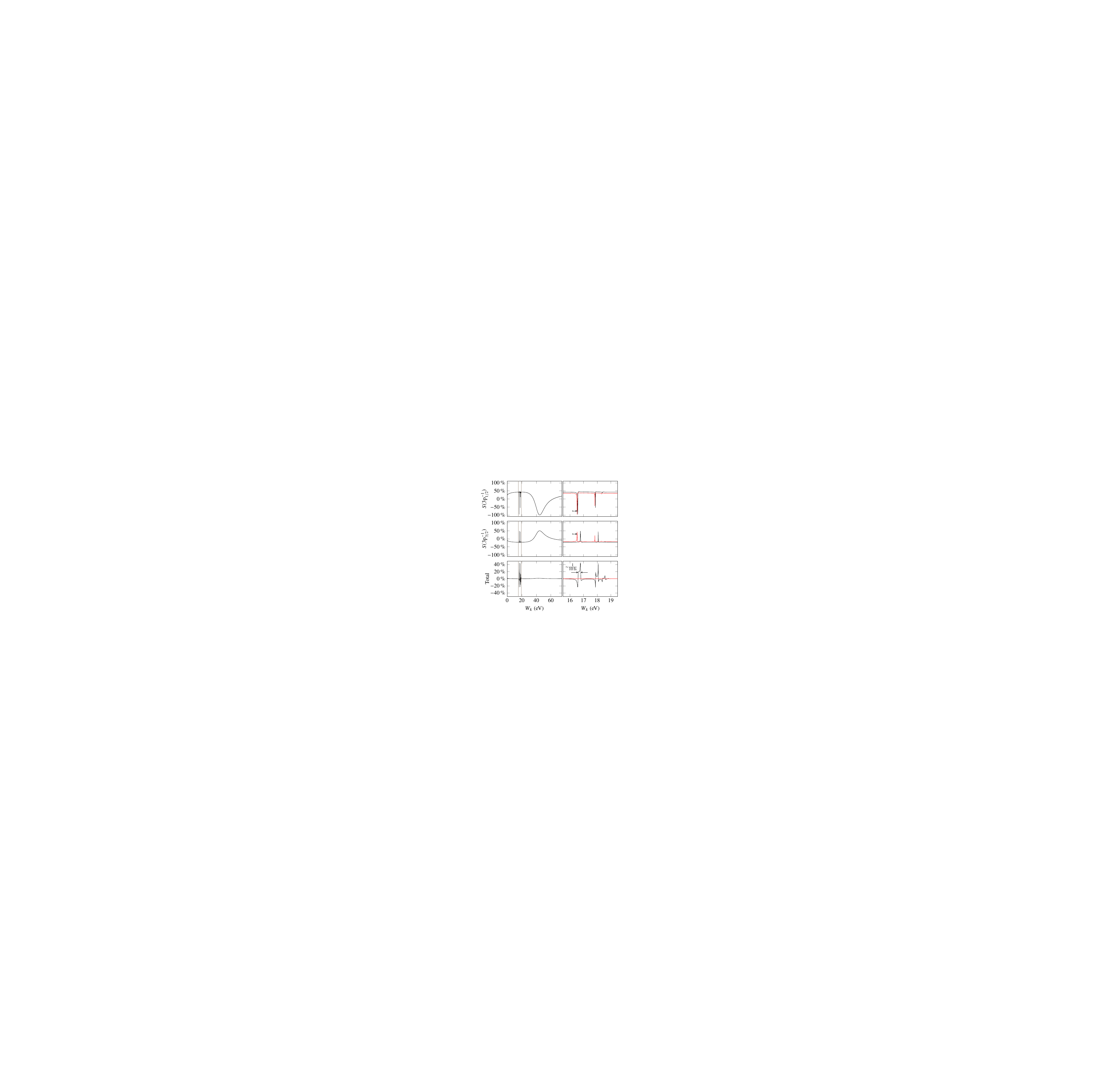}
  \caption{Argon angle-integrated photoelectron spin polarization for
    LCP light, resolved on ion energy level, i.e.\ which hole and
    \(J\), but not \(M_J\), as well as total. The \emph{left column}
    shows the entire energy range, including the CM. The \emph{right
      column} focuses on the Fano resonances leading up to the the
    \archannelc{} threshold, in the energy range delimited by the
    vertical lines in the left column. The black lines result from a
    calculation with ECP (i.e.\ quasirelativistic, including
    spin--orbit coupling). It is clearly visible that the Fano lines
    for \(J=1/2\) are shifted with respect to those for \(J=3/2\), a
    necessary condition for the generation of total spin
    polarization. The red lines, on the other hand, are the result of
    an all-electron, non-relativistic calculation; these have been
    blue-shifted by \SI{0.33}{\electronvolt} (as indicated by the
    arrow) to align the \archannelc{} threshold with the ECP
    calculation. In this case, the Fano lines for \(J=1/2\) and
    \(J=3/2\) appear at exactly the same energies, and therefore the
    spin polarization perfectly cancels when summing over all ion
    channels.}
  \label{fig:argon-ion-and-total-spin-pol}
\end{figure}

Once we average over all ionic states, the spin polarization is
substantially reduced, reaching a meagre \SI{2}{\percent} in the
vicinity of the Cooper minimum, see
Fig.~\ref{fig:argon-total-spin-pol}. This is in contrast to the case
of cæsium \cite{Fano1969,Heinzmann1970}, where around
\SI{85}{\percent} spin polarization was demonstrated, even when
angularly integrating the photoelectron distribution. Naturally,
overall spin polarization is absent for LP. For the Fano resonances,
the spin polarization remains high (reaching up to \SI{40}{\percent}).

What makes the CM of argon (and xenon) different from that of cæsium,
and why are the Fano resonances an efficient source of spin
polarization?  For argon (and xenon), where we ionize from
\(n\conf{p}^6\;\term{1}{S}[0]\) into
\(n\conf{p}^5\;(\term*{2}{P}[J])\;k\ell\), \(J=3/2,1/2\),
\(\ell=\conf{s},\conf{d}\), we have the following contributions:
\begin{equation}
  P_{J\spin}
  =
  \abs*{A_{J\spin}}^2
  \defd
  \abs*{\matrixel*{k\ell_{\spin}'m_\ell'}{\mp r\sphtensor[1][\pm1]}{n\conf{p}_Jm_J}}^2,
\end{equation}
where the upper (lower) sign corresponds to RCP (LCP). Evaluating the
spin--angular integrals (see the SI \citetheSI{}), summing over
\(\ell'\), \(m_\ell'\), and \(m_J\), and denoting the radial integrals as
\(R_{\ell',J}\), we find
\begin{equation}
  \bmat{P_{3/2\spinup} \\
    P_{3/2\spindown} \\
    P_{1/2\spinup} \\
    P_{1/2\spindown}}
  =
  \mat{C}_{\textrm{pol}}
  \bmat{\abs{R_{\conf{s},3/2}}^2 \\
    \abs{R_{\conf{d},3/2}}^2 \\
    \abs{R_{\conf{s},1/2}}^2 \\
    \abs{R_{\conf{d},1/2}}^2},
\end{equation}
\begin{equation}
  \begin{aligned}
    \mat{C}_{\textrm{RCP}}
    &=
      \frac{1}{9}
      \bmat{
        \tikzmarknode{R11}{1} & 5 & 0 & 0 \\
        \tikzmarknode{R21}{3} & 3 & 0 & 0 \\
        0 & 0 & \tikzmarknode{R33}{2} & 1 \\
        0 & 0 & \tikzmarknode{R43}{0} & 3
      },
    &
    \mat{C}_{\textrm{LCP}}
    &=
      \frac{1}{9}
      \bmat{
      \tikzmarknode{L11}{3} & 3 & 0 & 0\\
      \tikzmarknode{L21}{1} & 5 & 0 & 0\\
      0 & 0 & \tikzmarknode{L33}{0} & 3\\
      0 & 0 & \tikzmarknode{L43}{2} & 1},
  \end{aligned}
\end{equation}
\begin{tikzpicture}[remember picture, overlay]
  \draw[color1] ($ (R11.north west) + (-1pt,1pt) $) rectangle
  ($ (R21.south east)  + (1pt,-1pt) $);
  \draw[color2] ($ (R33.north west) + (-1pt,1pt) $) rectangle
  ($ (R43.south east)  + (1pt,-1pt) $);
  \draw[color1] ($ (L11.north west) + (-1pt,1pt) $) rectangle
  ($ (L21.south east)  + (1pt,-1pt) $);
  \draw[color2] ($ (L33.north west) + (-1pt,1pt) $) rectangle
  ($ (L43.south east)  + (1pt,-1pt) $);
\end{tikzpicture}
where the upper-left quadrant pertains to the \(J=3/2\) channel, and
the lower-right to the \(J=1/2\) channel. From this, we see that for a
\(J=1/2\) ion, we expect \SI{100}{\percent} spin-up (spin-down)
photoelectron for RCP (LCP) at the CM, where the radial integral
\(R_{\conf{d},1/2}\) vanishes:
\begin{equation}
  S =
  \pm\frac{2-0}{2+0} =
  \pm1.
\end{equation}
For a \(J=3/2\) ion, again at the CM where this time
\(R_{\conf{d},3/2}\) vanishes, we only expect partial spin
polarization:
\begin{equation}
  S =
  \pm\frac{1-3}{1+3} =
  \mp\frac{1}{2},
\end{equation}
consistent with the results observed in
Fig.~\ref{fig:argon-ion-and-total-spin-pol}. If we instead do not
observe the ion, again assuming that \(R_{\conf{d},J}\) vanish, we
expect the spin polarization to be
\begin{equation}
  S =
  \pm\frac{1+2-3}{1+2+3} =
  0.
\end{equation}
The reason why we observe any spin polarization at all is because the
Cooper minima for \(J=1/2\) and \(J=3/2\) are not perfectly aligned,
i.e.\ \(R_{\conf{d},1/2}\) and \(R_{\conf{d},3/2}\) do not vanish
simultaneously. Additionally, \(R_{\conf{s},1/2}\) and
\(R_{\conf{s},3/2}\) may differ slightly.

However, the presence of Fano resonances in argon with very rapid
variation of the dipole matrix elements
\cite{Starace1977,Meulen1991,Sorensen1994PRA,Carette2013} more than
makes up for this deficiency; the spectral feature around
\SI{16.7}{\electronvolt} in
Fig.~\ref{fig:argon-ion-and-total-spin-pol} is approximately
\SI{0.2}{\electronvolt} wide. This implies that ionizing with an LCP
pulse centred at \SI{16.7}{\electronvolt} that is about
\SI{10}{\femto\second} long, will yield photoelectrons with spin
polarization in excess of \SI{20}{\percent}, even when
\emph{integrating over all angles and not measuring the ion}; using a
longer pulse results in spin polarization up to \SI{40}{\percent}.

\begin{figure}
  \centering
  \includegraphics[trim=30cm 30cm 30cm 30cm]{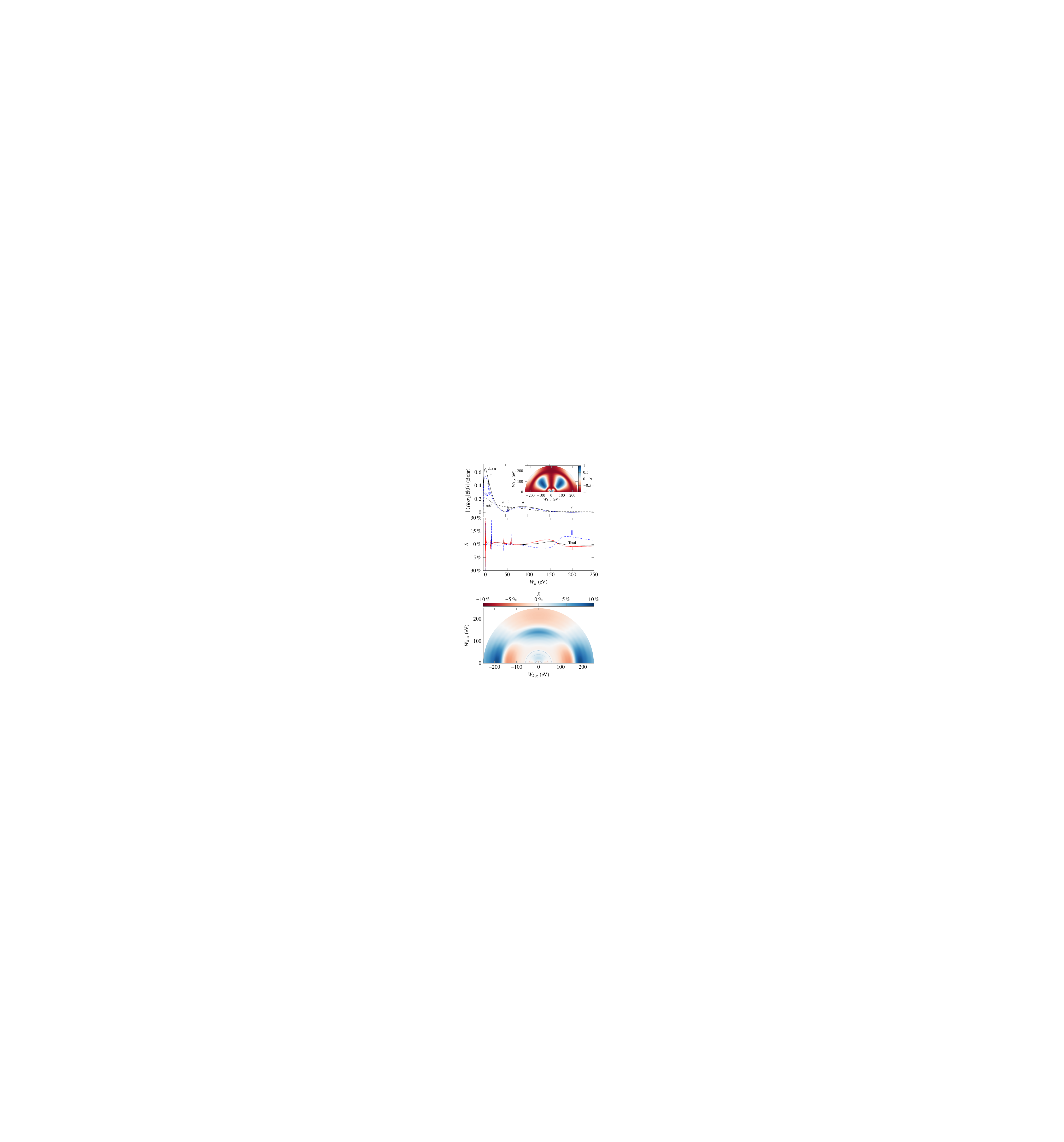}
  \caption{\emph{Top panel}: Xenon photoionization dipole matrix
    elements in the ionization channel \xechannelb{}, \(M_J=-1/2\),
    for the dipole moment along the \(z\) axis (see the SI
    \citetheSI{} for the very similar dipole moments for LCP and
    RCP). The different lines correspond to contributions from
    different partial waves and spins (solid spin-up and dashed
    spin-down). We observe two sets of Fano resonances, the first
    (indicated by \emph{a}) are due to the Rydberg states leading up
    to the \xechannelc{} threshold at around
    \(\kineng=\SI{16}{\electronvolt}\), and similarly, the second
    (\emph{c}) are due to the \xechanneld{} threshold at around
    \SI{59}{\electronvolt}. Additionally, the correlation-induced
    \cite{Kutzner1989} CM (\emph{b}) around \SI{50}{\electronvolt},
    the giant dipole resonance (\emph{d}) around
    \SI{100}{\electronvolt}, and a CM-like feature (\emph{e}) around
    \SI{200}{\electronvolt} are clearly visible.  \emph{Middle panel}:
    Xenon total spin polarization, i.e.\ obtained by summing over all
    ion channels, when ionizing with LCP light (RCP will give exactly
    the opposite result, and LP will yield exactly zero spin
    polarization), in the vicinity of the Fano resonances, the Cooper
    minimum, and the giant dipole resonance. The solid black line
    corresponds to angle-integrated spin-polarization; the dashed blue
    corresponds to emission along the forward direction (the backward
    direction is by symmetry exactly the same); the dotted red line
    corresponds to emission perpendicular to the polarization
    direction. \emph{Bottom panel}: Total spin polarization from
    xenon, resolved on the emission angle.}
  \label{fig:xenon-channel-dmes-and-total-spin-pol}
\end{figure}

We may perform the same analysis for xenon. There the situation is
more complex, due to the higher number of channels and the giant
dipole resonance present at around
\(\kineng=\SI{100}{\electronvolt}\), which leads to additional
structures in the dipole matrix elements, illustrated for the
\xechannelb{}, \(M_J=-1/2\) in
Fig.~\ref{fig:xenon-channel-dmes-and-total-spin-pol}. Again, we
predict perfect spin polarization in the vicinity of the two CM
(around \SI{50}{\electronvolt} and \SI{200}{\electronvolt},
respectively), this time \SI{100}{\percent} \emph{spin-down}, since we
consider the \(M_J=-1/2\) channel, i.e.\ the \enquote{mirror case} of
\(M_J=1/2\).  Again, when considering the total spin polarization (see
Fig.~\ref{fig:xenon-channel-dmes-and-total-spin-pol}), i.e.\ when
averaging over the ionic states, the spin polarization is rather small
in the vicinity of the CM and the giant dipole resonance, due the
presence of many more channels than in cæsium. Again, the spin
polarization is rather sizeable in the vicinity of the Fano resonances
(of which there are more than in argon), exceeding \SI{30}{\percent},
for some energies. In contrast to argon, the total spin polarization
depends strongly on the photoelectron emission angle, with opposite
sign along and perpendicular to the polarization direction.

\begin{acknowledgments}
  The work of SC has been supported through scholarship 185-608 from
  \emph{Olle Engkvists Stiftelse}. JMD acknowledges support from the
  \emph{Knut and Alice Wallenberg Foundation} (2017.0104 and
  2019.0154), the Swedish Research Council (2018-03845) and \emph{Olle
    Engkvists Stiftelse} (194-0734). MI acknowledges support from
  \emph{FET-OPEN \enquote{OPTOlogic}} (899794). OS acknowledges
  support from \emph{Horizon Europe} ERC-2021-ADG
  (\href{https://doi.org/10.3030/101054696}{101054696 Ulisses}).
\end{acknowledgments}

\bibliographystyle{apsrev4-2}
\bibliography{\bibliographyfile}

\clearpage

\appendix

\section*{Supplementary information}
\label{app:supplementary-information}

\setcounter{figure}{0}
\renewcommand\thefigure{SI--\arabic{figure}}
\setcounter{equation}{0}
\renewcommand\theequation{SI--\arabic{equation}}
\renewcommand\thesubsection{\Alph{subsection}}

\subsection*{Electronic structure}
\begin{figure}[htb]
  \centering
  \includegraphics[trim=30cm 30cm 30cm 30cm]{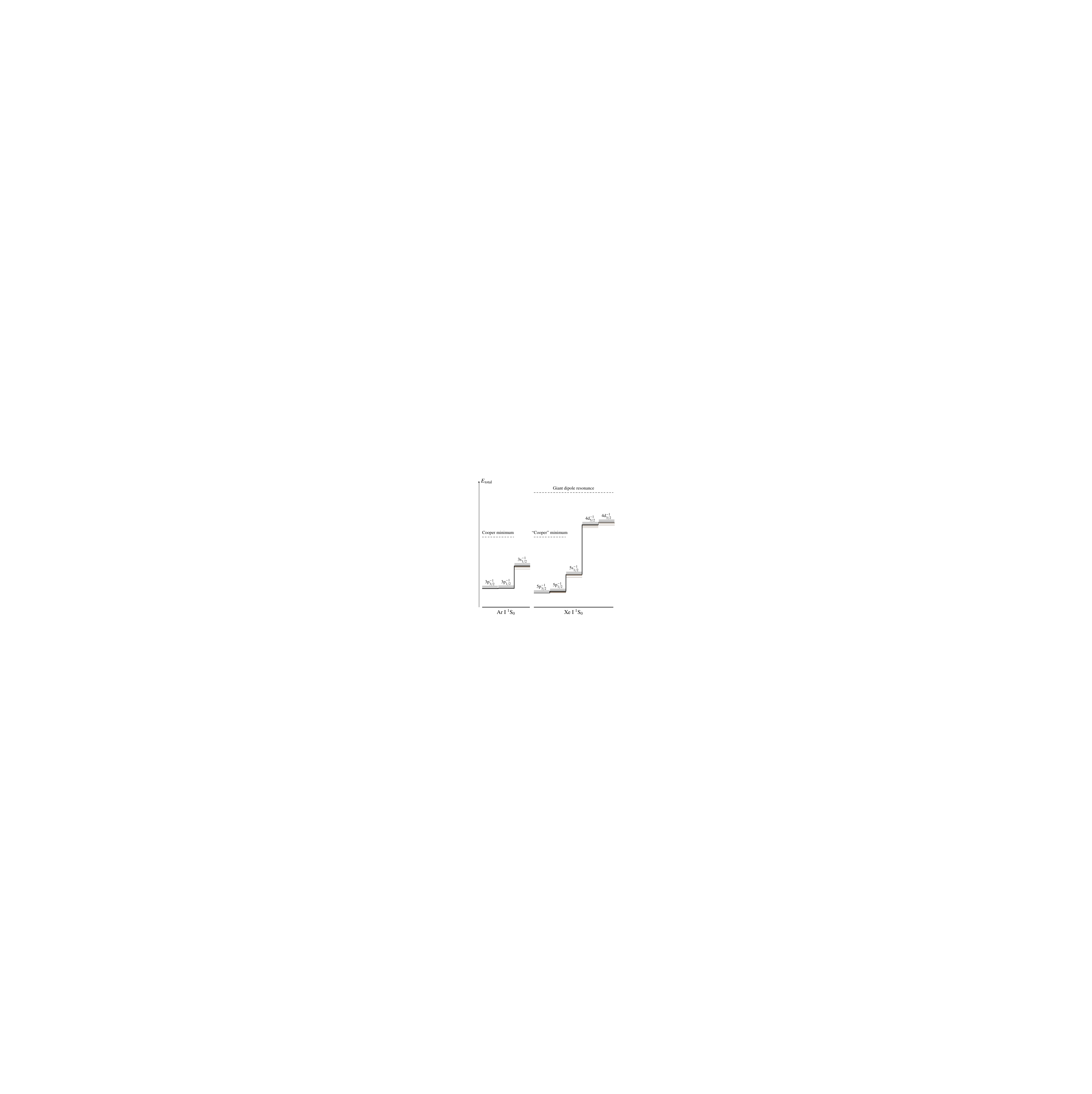}
  \caption{Sketch of the energy structures of argon and xenon; the
    dashed line indicates the position of the CM in the \archannela{}
    and \archannelb{} channels in argon. Also in xenon, there is a
    minimum in the matrix elements just before the giant resonance,
    which is similar to a CM, but is correlation-induced due to the
    interaction with the \conf{4d} electrons \cite{Kutzner1989}. The
    sketch is to scale with energies as listed in
    Table~\ref{tab:ionization-energies}. Also shown are the
    autoionizing states in each channel that will lead to Fano
    resonances in all channels shown to the left of them (i.e.\ with
    lower \(\ionpotential\)).}
  \label{fig:energy-diagram}
\end{figure}
\begin{table}[htb]
  \caption{\label{tab:ionization-energies} Theoretical ionization
    potentials of the \(n\conf{s}\) and \(n\conf{p}\) electrons of
    argon and xenon, compared with their experimental and relativistic
    theoretical values (Refs: \citeSI{Velchev1999}\dataref{a},
    \citeSI{Saloman2010}\dataref{b}, \citeSI{Saloman2004}\dataref{c},
    \citeSI{Hansen1987PS}\dataref{d}, \citeSI{Codling1964}\dataref{e},
    \citeSI{Cardona1978-back-matter}\dataref{f},
    \citeSI{Bearden1967}\dataref{g}, \cite{Zapata2022}\dataref{h}).}
  \myruledtabular{ll|S[table-format=2.3]S[table-format=2.4]S[table-format=3.4]S[table-format=3.1]S[table-format=3.4]}{%
    Element & Hole & \mc{\unitlabel{\(\ionpotential\)}{\electronvolt}} & \mc{Exp.\ \unitbracket{\electronvolt}} &
    \mc{\unitlabel{\(\Delta\)}{\electronvolt}} &
    \mc{\unitlabel{Rel.\(\Delta\)}{\percent}} &
    \mc{\unitlabel{\(\ionpotential\)}{\electronvolt}}
    \Bstrut\\
    \hline
    \Tstrut
    Ar & \archannela & 15.871 & 15.760\dataref{a} & 0.111 & 0.7 & 15.995\dataref{h} \\
    \Tstrut
    & \archannelb & 16.069 & 15.937\dataref{b} & 0.132 & 0.8 & 16.201\dataref{h} \\
    \Tstrut
    & \archannelc & 35.159 & 29.239\dataref{b} & 5.920 & 20.2 & 35.010\dataref{h} \\
    \hline
    \Tstrut
    Xe & \xechannela & 12.026 & 12.130\dataref{c} & -0.104 & -0.9 & 11.965\dataref{h} \\
    \Tstrut
    & \xechannelb & 13.483 & 13.436\dataref{d} & 0.047 & 0.3 & 13.407\dataref{h} \\
    \Tstrut
    & \xechannelc & 27.927 & 23.397\dataref{d} & 4.530 & 19.4 & 27.507\dataref{h} \\
    \Tstrut
    & \xechanneld & 70.481 & 67.55\dataref{e} & 2.98 & 4.4 & 71.640\dataref{h} \\
    \Tstrut
    & \xechannele & 72.591 & 69.52\dataref{e} & 3.09 & 4.4 & 73.760\dataref{h} \\
    \Tstrut
    & \xechannelf & 162.450 & 145.5\dataref{f} & 16.950 & 11.6 & 162.801\dataref{h} \\
    \Tstrut
    & \xechannelg & 175.522 & 146.7\dataref{g} & 28.822 & 19.6 & 175.610\dataref{h} \\
    \Tstrut
    & \xechannelh & 228.995 & 213.2\dataref{f} & 15.795 & 7.4 & 229.474\dataref{h} }
\end{table}
The ground state \(\ket{0}\) is approximated using the Hartree–Fock
solution. Since we wish to consider the spin–orbit interaction, the
orbitals are expanded in two-component spinor spherical harmonics (see
section 7.2 of \cite{Varshalovich1988}) which are eigenfunctions of
\(\operator{L}^2\), \(\operator{S}^2\), \(\operator{J}^2\), and
\(\operator{J}_z\). The spin–orbit interaction is thus diagonal in
this basis, and is implemented using effective core potentials (ECPs),
which account for scalar- and vector-relativistic effects. For argon,
we use the large-core ECP by \textcite{Nicklass1995}, and for xenon
the small-core ECP by \textcite{Peterson2003}. The ion states
\(\ket{I}\) are approximated at the configuration-interaction singles
(CIS) level, which corresponds to removal of one electron, with the
remaining orbitals frozen, i.e.\ no relaxation. This leads to slightly
too high ionization potentials, especially for inner electrons, as
seen in Table~\ref{tab:ionization-energies}. In the calculations
presented in this work, ionization was only permitted from
\conf{3s}--\conf{3p} in case of argon, and from \conf{4d}--\conf{5p} in
case of xenon (see Fig.~\ref{fig:energy-diagram}).

\subsection*{Dipole matrix elements}

\begin{figure}
  \centering
  \includegraphics[trim=30cm 30cm 30cm 30cm]{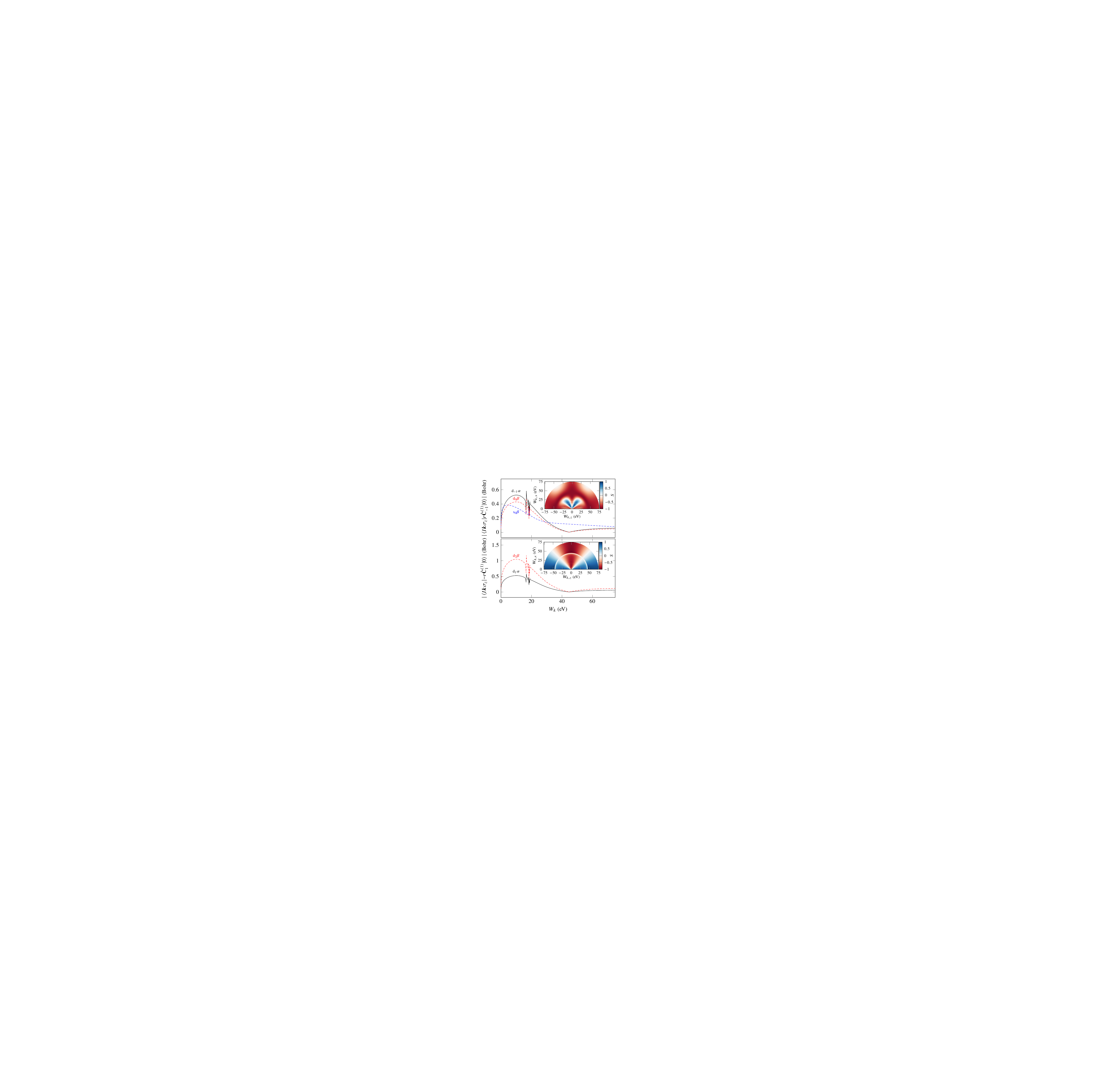}
  \caption{Argon photoionization dipole matrix elements in the
    ionization channel \archannelb{}, \(M_J=1/2\), for the dipole
    moments for LCP (\emph{top panel}), and RCP (\emph{bottom
      panel}). These are formed from linear combinations of
    \(\operator{x}\) and \(\operator{y}\) (see main text). The
    different lines correspond to contributions from different partial
    waves and spins (solid spin-up and dashed spin-down).}
  \label{fig:argon-channel-dmes-xy}
\end{figure}

\begin{figure}
  \centering
  \includegraphics[trim=30cm 30cm 30cm 30cm]{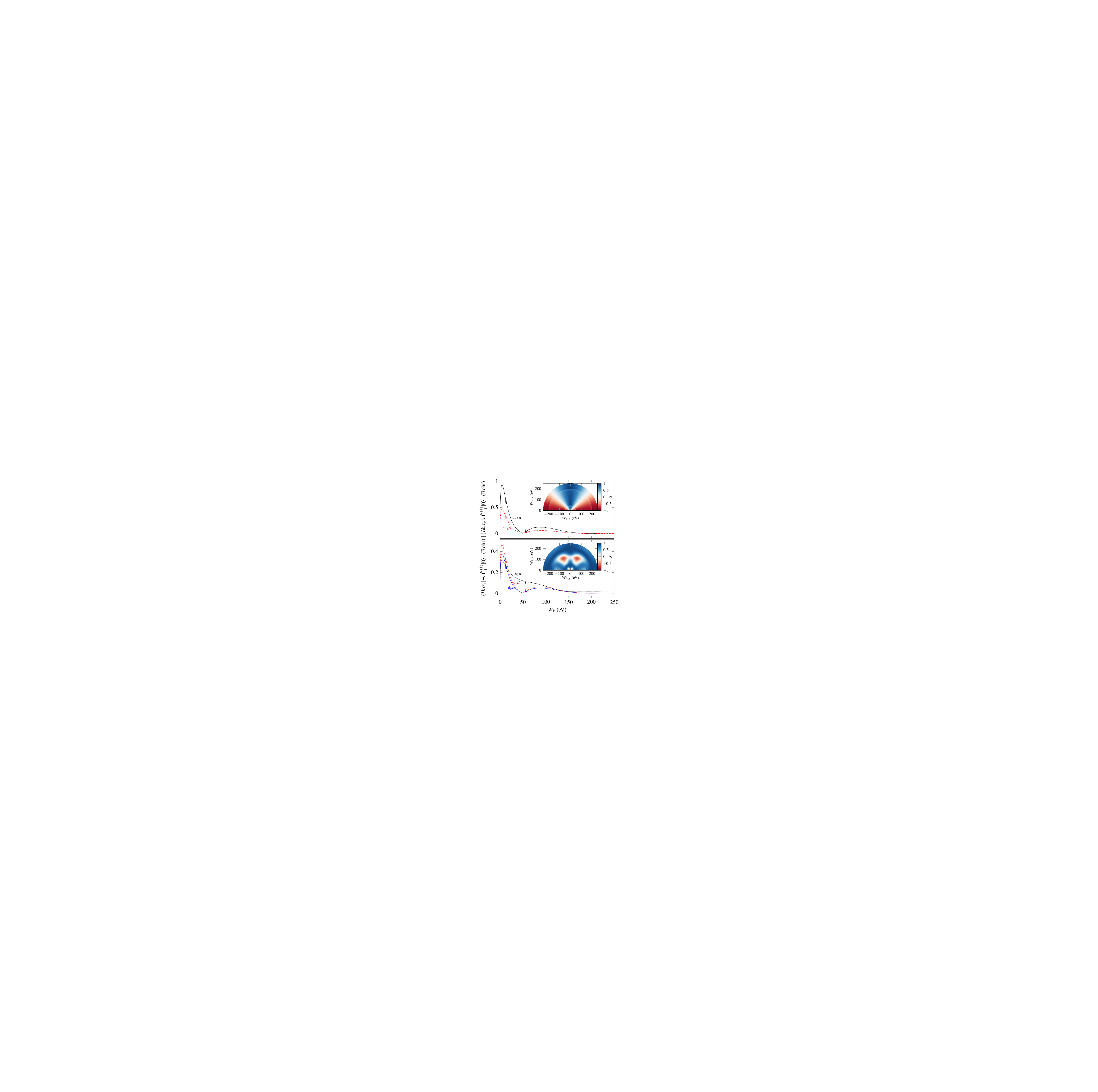}
  \caption{Xenon photoionization dipole matrix elements in the
    ionization channel \xechannelb{}, \(M_J=-1/2\), for the dipole
    moments for LCP (\emph{top panel}), and RCP (\emph{bottom
      panel}). The different lines correspond to contributions from
    different partial waves and spins (solid spin-up and dashed
    spin-down).}
  \label{fig:xenon-channel-dmes-xy}
\end{figure}

\begin{figure}
  \centering
  \includegraphics[trim=30cm 30cm 30cm 30cm]{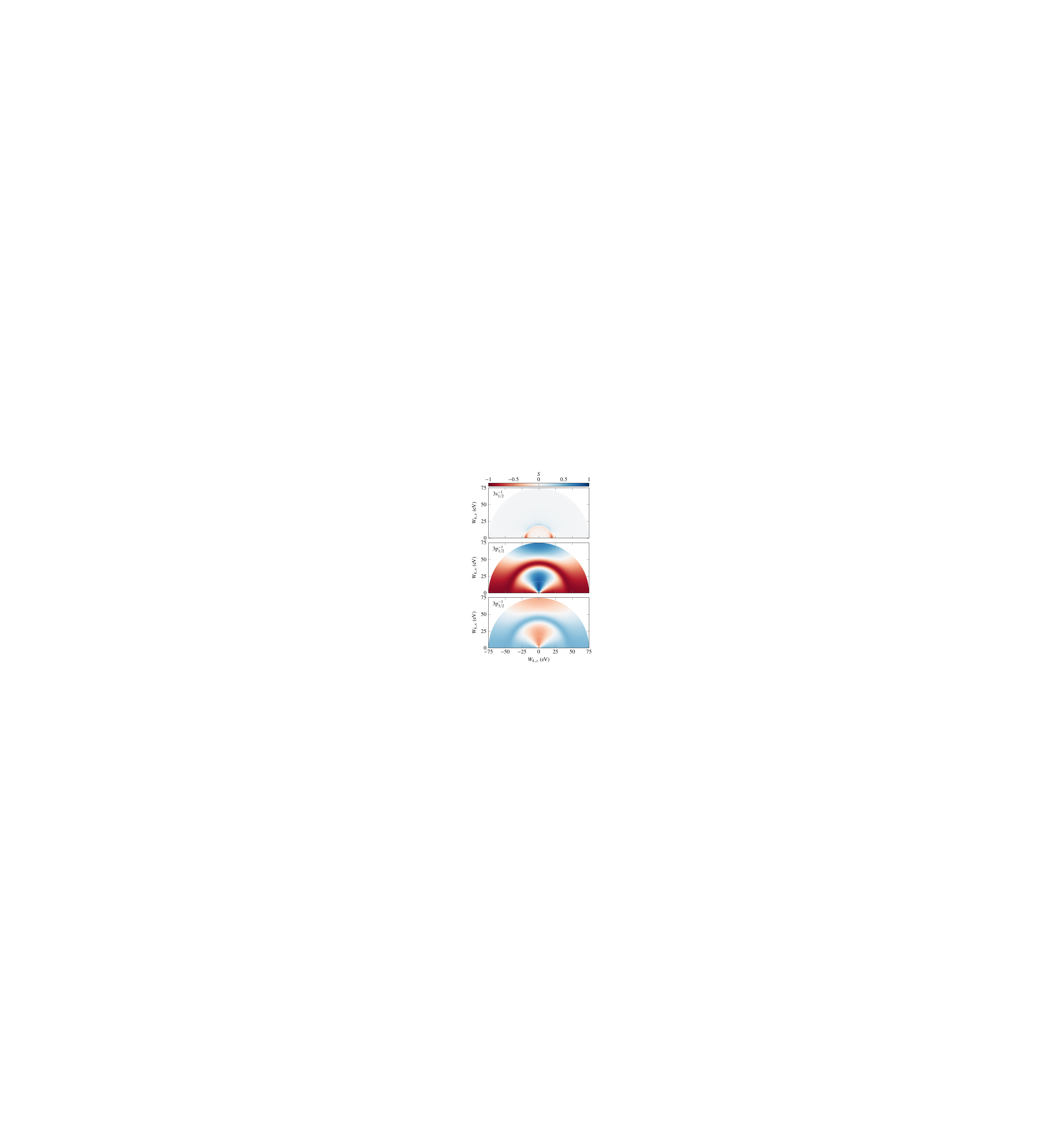}
  \caption{Ion-resolved (i.e.\ \(J\)-resolved) spin polarization for
    LCP ionization (RCP will give exactly the opposite result, and LP
    will yield exactly zero spin polarization) of argon.}
  \label{fig:argon-ion-spin-pol-angle-resolved}
\end{figure}

\begin{figure}
  \centering
  \includegraphics[trim=30cm 30cm 30cm 30cm]{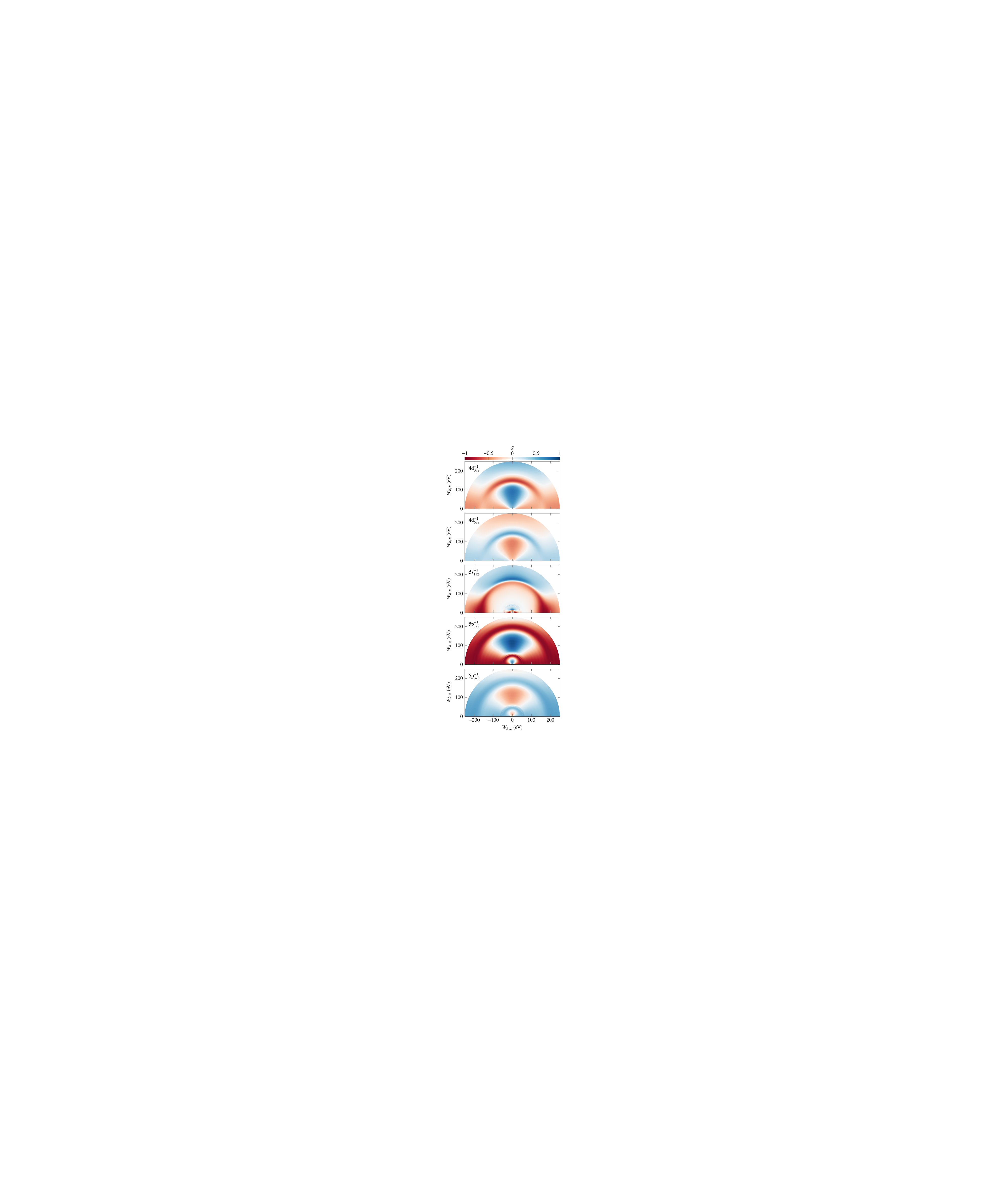}
  \caption{Ion-resolved (i.e.\ \(J\)-resolved) spin polarization for
    LCP ionization (RCP will give exactly the opposite result, and LP
    will yield exactly zero spin polarization) of xenon.}
  \label{fig:xenon-ion-spin-pol-angle-resolved}
\end{figure}

The dipole moments \(\operator{x}\), \(\operator{y}\), and
\(\operator{z}\) can be written in terms of the components of the
spherical tensor \(\sphtensor[1]\) as
\begin{equation}
  \begin{aligned}
    \operator{x}
    &=
      \frac{r}{\sqrt{2}}
      [-\sphtensor[1][1] + \sphtensor[1][-1]],
    &
      \operator{y}
    &=
      \frac{r}{\sqrt{2}}
      [\im\sphtensor[1][1] + \im\sphtensor[1][-1]],
    &
      \operator{z}
    &= r\sphtensor[1][0].
  \end{aligned}
\end{equation}
\(\operator{x}\) and \(\operator{y}\) thus have precisely the same
angular structure, i.e.\ they both couple to
\(m_\ell' = m_\ell \pm 1\), but with different complex phases. The difference
in phase between the matrix element of \(\operator{x}\) and
\(\operator{y}\) is what leads to the asymmetry in spin polarization
between LCP
[\((\operator{x}-\im\operator{y})/\sqrt{2} \equiv r\sphtensor[1][-1]\)] and
RCP
[\((\operator{x}+\im\operator{y})/\sqrt{2} \equiv -r\sphtensor[1][1]\)];
see Fig.~\ref{fig:argon-channel-dmes-xy} for argon and
Fig.~\ref{fig:xenon-channel-dmes-xy} for xenon. Additionally,
\(J\)-resolved (i.e.\ traced over \(M_J\)) spin polarizations are
shown in Fig.~\ref{fig:argon-ion-spin-pol-angle-resolved} and
Fig.~\ref{fig:xenon-ion-spin-pol-angle-resolved} for argon and xenon,
respectively.

The explicit formula for the ion- and electron-spin-resolved dipole
matrix elements from the ground state \(n\conf{p}^6\;\term{1}{S}[0]\)
is given by
\begin{equation}
  \begin{aligned}
    A_{J\spin}
    &=
      \matrixel*{k\ell_{\spin}'m_\ell'}{\mp r\sphtensor[1][\pm1]}{n\conf{p}_Jm_J} \\
    &=
      \mp
      R_{\ell',J}
      \matrixel*{\ell' m_\ell'}{\sphtensor[1][\pm1]}{\conf{p}m_\ell}
      \CG*{\conf{p},m_\ell}{\frac{1}{2},\spin}{J, m_J} \\
    &=
      \mp
      [2\ell'+1]^{-1}
      R_{\ell',J}
      \CG*{\conf{p},m_\ell}{1,\pm1}{\ell', m_\ell'}
      \matrixelred*{\ell'}{\sphtensor[1]}{\conf{p}}
      \CG*{\conf{p},m_\ell}{\frac{1}{2},\spin}{J, m_J} \\
    &=
      \mp
      \sqrt{3}
      [2\ell'+1]^{-1}
      R_{\ell',J}
      \CG*{\conf{p},m_\ell}{1,\pm1}{\ell', m_\ell'}
      \CG*{\conf{p},0}{1,0}{\ell',0}
      \CG*{\conf{p},m_\ell}{\frac{1}{2},\spin}{J, m_J},
  \end{aligned}
\end{equation}
where
\begin{equation}
  R_{\ell',J}
  \defd
  \int_0^\infty
  \diff{r}
  P(k\ell';r)
  r
  P(n\conf{p}_J;r).
\end{equation}

\subsection*{Technical details on computing spin polarization}
Computing the spin polarization according to
Eq.~\eqref{eqn:spin-polarization-matrix-elements} of the main article,
care is needed when \(P_\spinup\) and \(P_\spindown\) vanish
simultaneously. Otherwise, due to numerical round-off errors, one
might divide something small by zero, and the computed spin
polarization will then exhibit spikes. One way of circumventing this
is to add a regulator to the denominator, or better a small fraction
of a smoothed copy of the denominator itself:
\begin{equation}
  \label{eqn:mollified-spin-polarization}
  S \defd
  \frac{P_\alpha-P_\beta}{\Xi(P_\alpha+P_\beta)},
  \quad
  \Xi(y) = y + \epsilon\average{y},
\end{equation}
where \(\average{y}\) is a smooth version of \(y\), and \(\epsilon\) a
scaling factor.

When the electron is measured in coincidence with the ion, resolved on
\(J\) and \(M_J\), the coordinate system of the ion is \emph{fully}
specified, i.e.\ not only the quantization axis \(z\), which is set by
the external field, but also \(x\) and \(y\). Since these are
typically not fixed in the experiment, they must be averaged over; it
is not enough to sum over \(M_J\). If this is not done, a preferred
orientation of the universe is chosen, and spin polarization dependent
on the photoelectron emission angle \(\phi\) will be observed.

\subsection*{Description of RTDCIS calculations}

The electronic structure described above is only quasirelativistic,
and we use the RTDCIS \cite{Zapata2022,Tahouri2024} calculations as an
independent, fully-relativistic benchmark of our perturbation theory
results. These calculations are however limited to LP at the
moment. They provide the particle--hole channel-resolved photoelectron
populations characterized by the quantum numbers \(\ell\), \(j\), and
\(m_j\). Given a relativistic hole \(a\) in the ion, we define a
one-electron time-dependent orbital as:
\(\ket*{ \chi_{a}(t)}=\sum_p c^p_{a}(t) \ket*{ \phi_{p}}\) where
\(\ket*{ \phi_{p}}\) are virtual electron orbitals with energies
\(\epsilon_p\), and \(c^p_{a}(t)\) are their time-dependent complex
amplitudes \citeSI{Rohringer2006}. For a given particle--hole channel,
the final photoelectron population is given by
\({P^{\ell jm_j}_a}=\lim_{t \rightarrow \infty} \sum_p \vert c_a^p(t)\vert^2\). We obtain the
absolute value of the resonant complex amplitude as
\(\vert c^{\ell jm_j}_{a} \vert = (P^{\ell jm_j}_a)^{1/2}\) with
\(\epsilon_p \approx \epsilon_a + \omega\), where the
\(P^{\ell jm_j}_a\) is computed using RTDCIS, and \(\epsilon_a\) is the energy
of the occupied orbital \(\ket{\phi_a}\).

The \(\ell\)-, \(m_\ell\)- and spin-resolved probabilities associated with a
given hole \(a\) are given by:
\begin{equation}
  \begin{aligned}
    P_{a}^{\ell m_\ell m_s}
    &=
    \lim_{ t \rightarrow \infty}
    \braket*{\chi_{a}(t)}{\ell m_\ell m_s}
    \braket*{\ell m_\ell m_s}{\chi_{a}(t)}, \\
    c_a^{\ell m_\ell m_s}(t)
    &\defd
      \braket*{\ell m_\ell m_s}{\chi_{a}(t)} =
      \braket*{\ell m_\ell m_s}{\ell j m_j}
      \braket*{\ell j m_j}{\chi_{a}(t)} \\
    &=
      \braket*{\ell m_\ell m_s}{\ell j m_j}
      c_a^{\ell j m_j}(t),
  \end{aligned}
\end{equation}
where we implicitly sum over \(j\) and \(m_j\). If there is only one
\(j\) involved, the spin probabilities \(P_{a}^{\ell m_\ell m_s}\) can be
expressed as a function of the absolute values of the complex
amplitudes, \(\abs*{c^{\ell jm_j}_{a}}\). When there are contributions
from two states with same \(\ell\) but different total angular momenta,
\(j_1\neq j_2\), there are also cross terms involving the phase
difference between the two complex amplitudes. This phase difference
can be obtained by the ratio
\(\phi=\arg\{c^{\ell{}j_1m_j}_a/c^{\ell{}j_2m_j}_a\}\) \citeSI{Dahlstroem2019}.
According to perturbation theory, for LP these complex amplitudes are
proportional to the matrix elements of the dipole operator $\hat{z}$:
\(c^{\ell jm_j}_a \propto \matrixel*{\epsilon_p \ell j m_j}{\operator{z}}{n'\ell'j'm_j}\),
where \(\ket*{n'\ell'j'm_j}\) and \(\ket*{\epsilon_p \ell jm_j}\) represent the
hole and particle states, respectively. The complex amplitudes can be
estimated with the aid of the Wigner--Eckart theorem:
\begin{equation}
  \begin{aligned}
    c^{\ell jm_j}_a
    &\propto
      \matrixel*{ \epsilon_p \ell j m_j }{ \operator{z} }{ n' \ell' j' m_j } \\
    &= (-)^{j-m_j}
      \begin{pmatrix}
        j    & 1 & j'\\
        -m_j & 0 & m_j
      \end{pmatrix}
      \matrixel*{ \epsilon_p \ell j}{ r }{ n' \ell' j' }
      \matrixelred*{ j }{ \sphtensor[1] }{ j' }
  \end{aligned}
\end{equation}
where the reduced matrix element of the spherical tensors of rank
\(k\) for half-integer angular momenta is given by \citeSI{Grant2007}
\begin{equation}
  \matrixelred*{ j }{ \sphtensor[k] }{ j' } =
  (-)^{j+1/2} [(2j+1)(2j'+1)]^{1/2}
  \begin{pmatrix}
    j & k & j'\\
    1/2 & 0 & -1/2
  \end{pmatrix}.
\end{equation}

\clearpage

\bibliographystyleSI{apsrev4-2}
\bibliographySI{\bibliographyfile}

\vfill

\end{document}